\documentclass[twocolumn]{IEEEtran}

\usepackage{times,amsmath}
\usepackage{bm}
\usepackage{epsfig}
\usepackage{amsfonts}
\usepackage{subfigure}
\usepackage{color}
\usepackage{lipsum}
\usepackage{float}

\title{On Modeling Perfectly Conducting Sharp Corners With Magnetically Inert Dielectrics Of Extreme Complex Permittivities}

\author{\IEEEauthorblockN{Constantinos A. Valagiannopoulos and Ari Sihvola}\\
\IEEEauthorblockA{Department of Radio Science and Engineering,\\
                  School of Electrical Engineering, Aalto University, Finland,\\
                  PO Box 13000, FIN-00076 AALTO.\\
                  Email: \{konstantinos.valagiannopoulos, ari.sihvola\}@aalto.fi}}

\begin{document}
\maketitle

\begin{abstract}
The idea of replacing an edgy perfectly conducting boundary by the corresponding interface filled with a dielectric material of extreme complex permittivities, is examined in the present work. A semi-analytical solution to the corresponding boundary value problems is obtained and the merit of the modeling has been checked. Certain conclusions for the effect of the constituent material parameters and the geometric features of the configuration on the model effectiveness, are drawn and discussed.   
\end{abstract}


\section{Introduction}
Fictitious, non-penetrable boundaries such as perfect electrically conducting (PEC) surfaces are frequently used in electromagnetic problems to simplify the solution process. Impedance boundary conditions have been also extensively utilized for the same reason in electromagnetic modeling \cite{HoppeRahmatSamii}; in particular, an interesting historical review focusing on the philosophy of such a concept is given in \cite{ImpedanceConditions}. The condition of a perfect magnetic conductor (PMC) is a very widely-used concept in electromagnetic modeling, and several designs have been presented in the literature to fabricate artificial magnetic conductor surfaces \cite{Sievenpiper,KildalKishk}. Moreove, PEC and its dual, PMC, can also be seen as limiting cases of a more general ideal boundary: the perfect electromagnetic conductor \cite{PerfectEMConductor}, which is employed for treating bianisotropic scattering through certain transformations. Still a further class of ideal boundaries is determined by forcing conditions on the normal components of the electric and magnetic flux densities (instead of the tangential components of the fields) which leads to the so-called DB boundary and its generalizations \cite{DB}. 

In the theoretical and computational analysis, when modeling antennas and other electromagnetic structures, these ideal boundary conditions help in confining the computational domain into the region of interest and hence reducing the required amount of computations. However, once the design has potential to be fabricated, these mathematical conceptions should be materialized properly in actual structures. The first step towards realization of such boundaries involves replacing the boundary by an interface with penetrable substances which usually needs to be of extreme permittivity and permeability as shown for the PEC case in \cite{ModelingPEC}. However, a full realization needs to proceed still further: first to find out what type of material microstructure reproduces the macroscopically extreme parameter values, and once the recipe has been found, finally to fabricate it using available materials, like metals and dielectric media.

Problems of scattering and diffraction by infinite metallic wedges have been thoroughly studied due to their canonical importance and suitability to analytical modeling. They are widely used to model certain configurations as such parts of corner reflectors, open-ended waveguides, transmission lines and other devices with polygonal shape existing in a laboratory of electromagnetics. A rigorous solution to the problem of two identical perfectly conducting parallel wedges loaded with a dielectric cylinder, is proposed in \cite{DoubleWedge}. In addition, wave diffraction by a wedge with anisotropic impedance boundary conditions has been explicitly solved through a probabilistic random walk method \cite{RandomWalk}. Finally, a sharp PEC corner with a cylindrical protecting cap is investigated in \cite{CappedCorner}, where possible applications are also presented.

In this work, we combine the two aforementioned topics (boundary modeling, sharp corners) to examine the  possibility of materializing a PEC wedge with use of lossy dielectric media. We consider a two-dimensional configuration constituted by two thin consecutive slabs (hands) excited by a plane wave. Due to the finite size of the construction, we perform a discretization of the hands into a large number of tiny cylinders and a semi-analytic solution is obtained. The actual current of the metallic case is compared to the polarization current of the dielectric case and similar comparisons have been made for the scattering field. Furthermore, the difference in the developed field close to the edge, is represented with respect to the real and imaginary permittivity of the dielectric, with several angular extents and hand lengths, for various slab thicknesses and excitation angles. In this sense, certain conclusions on the effectiveness of the PEC wedge modeling via magnetically inert materials are stated and explained.

\begin{figure}[t]
\centerline{\includegraphics[scale=0.5]{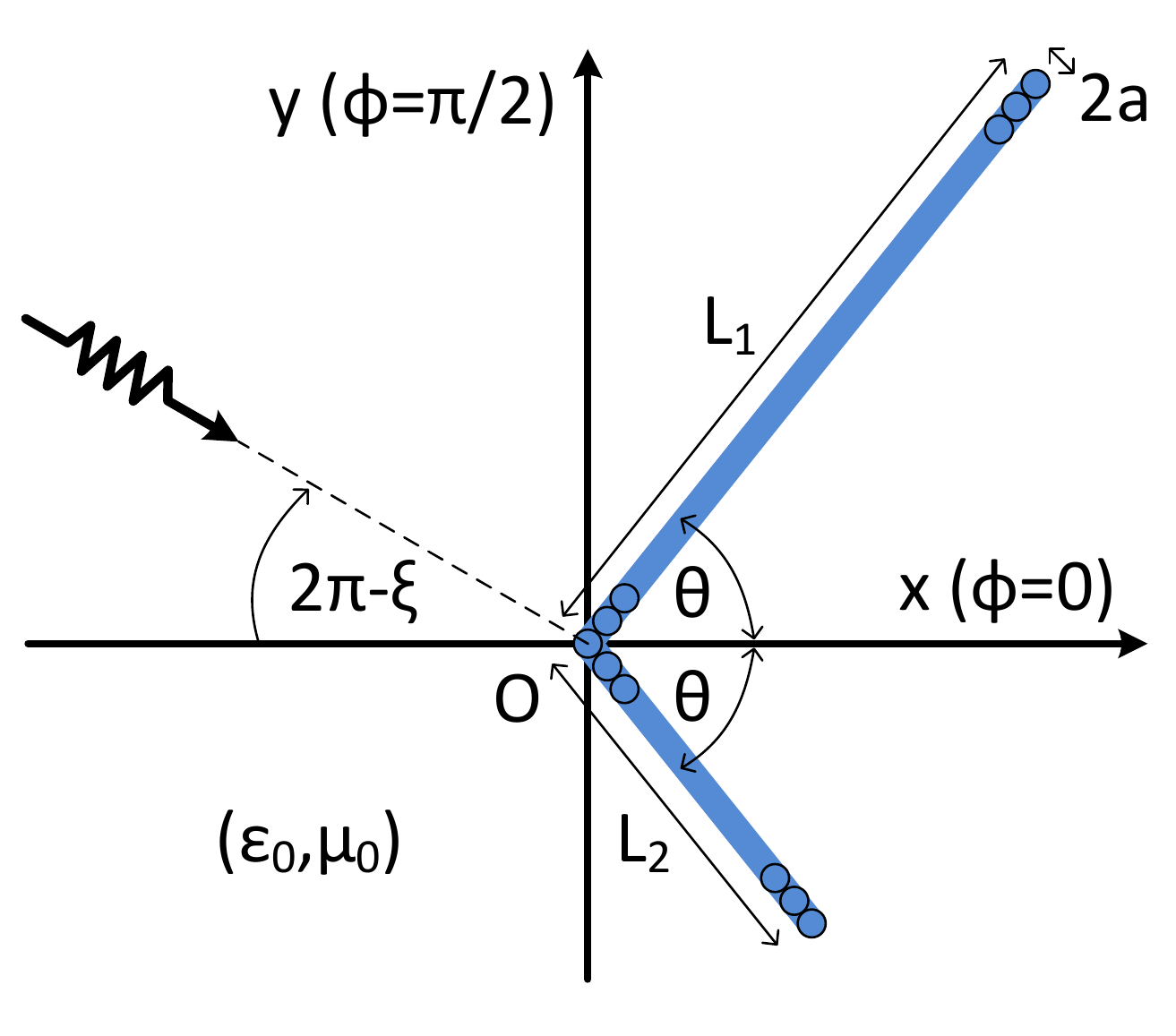}}
\caption{The configuration of the corner structure. The two thin consecutive slabs forming a wedge can be either metallic or constructed from magnetically inert dielectric material.} 
\label{fig:Figu1}
\end{figure}

\section{Mathematical Methodology}
\subsection{Primary Definitions}
In Fig. \ref{fig:Figu1}, the physical configuration of the considered construction is shown, and also the interchangeably used coordinate systems (Cartesian $(x,y,z)$ and cylindrical $(\rho,\phi,z)$) are defined. Two thin slabs, with different lengths $L_1,L_2$, are forming a corner of angular extent $2\theta$, placed symmetrically with respect to $x$ axis. The thickness of the strips is not infinitesimal, yet small, equaling to $2a<<L_1+L_2$. The whole two-dimensional structure is illuminated by a plane wave, of unitary magnitude, whose propagation direction makes an angle $\xi$ with the positive $x$ semi-axis. The described corner could be either metallic or from dielectric with intrinsic parameters $(\epsilon_r\epsilon_0, \mu_0)$. The scope of the present work is to study qualitatively and quantitatively the conditions under which the dielectric wedge, close to the sharp corner, responds as similar to the metallic analogous as possible. The time dependence is of harmonic type $e^{+j\omega t}$, where the operating circular frequency is denoted by $\omega$ and the free-space wavenumber by $k_0=\omega\sqrt{\epsilon_0\mu_0}=2\pi/\lambda_0$. 

Due to the nature of the configuration and the excitation given by: $\textbf{E}_{inc}(\rho,\phi)=\textbf{z}E_{inc}(\rho,\phi)=\textbf{z}e^{-jk_0\rho\cos(\phi-\xi)}$, the electric field in any region will be $\textbf{z}$-polarized $(\textbf{E}(\rho,\phi)=\textbf{z}E(\rho,\phi))$. The Green's function, equaling a line source response of electric current $j/(\omega\mu_0)$ positioned along the axis $(\rho,\phi)=(P,\Phi)$, is given by:
\begin{eqnarray}
G(\rho,\phi, P,\Phi)=-\frac{j}{4}\cdot \nonumber \\
\sum_{n=-\infty}^{+\infty}J_n(k_0\min(\rho,P))H_n^{(2)}(k_0\max(\rho,P))e^{-jn(\phi-\Phi)},
\label{eq:GreenFunction}
\end{eqnarray}
where $J_n,H_n^{(2)}$ are the Bessel and second-type Hankel functions of $n$-th order respectively. To render our boundary value problem susceptible to semi-analytical treatment, we regard the two hands of the strip wedge separated in a large number of consecutive thin cylinders with radius $a$. Two numberings of these pins ($u_q=1,\cdots, U_q=\lceil\frac{L_q}{2a}\rceil$ with $q=1,2$) are used in proportion to which hand the cylinders are referred to. The positions of the centers of the pins (in the cylindrical coordinate system) constituting the upper (first) and the lower (second) hand of the wedge, are given by:
\begin{subequations}
\begin{equation}
(\rho_{1}(u_1), \phi_1(u_1))=(2au_1,  \theta), ~u_1=1,\cdots, U_1, \\
\label{eq:PinsCentersPositions1}
\end{equation}
\begin{equation}
(\rho_{2}(u_2), \phi_2(u_2))=(2au_2, -\theta), ~u_2=1,\cdots, U_2.
\label{eq:PinsCentersPositions2}
\end{equation}
\label{eq:PinsCentersPositions}
\end{subequations}
Finally, the circular boundaries of the pins and the corresponding areal disks are denoted by $c_q(u_q)$ and $s_q(u_q)$ respectively ($q=1,2~,~u_q=1,\cdots,U_q$).

\subsection{Metallic Corner}
Let us first consider the metallic wedge. It is well-known that a rudimentary analytic tool for solving wave diffraction by PEC scatterers, is the radiation integral \cite{RadiationIntegral}. According to this important formula, the scattering field is given by (\ref{eq:MetallicIntegral}) which is appeared on the top of the next page. 
\begin{figure*}[t]
\hrulefill
\setcounter{equation}{2}
\begin{equation}
E_{scat}(\rho,\phi)= -j\omega\mu_0
\left[\sum_{u_1=1}^{U_1}k_1(u_1)\int_{c_1(u_1)}G(\rho,\phi,P_C,\Phi_C)dC+\sum_{u_2=1}^{U_2}k_2(u_2)\int_{c_2(u_2)}G(\rho,\phi,P_C,\Phi_C)dC\right].
\label{eq:MetallicIntegral}
\end{equation}
\end{figure*}
The notation $k_q(u_q)$ corresponds to the $\textbf{z}$-directed unknown line current flowing in the $u_q$-th cylinder. The length variable $C$ is used for the line integrations around the circular boundaries of the pins with respect to the arguments $(P_C, \Phi_C)$. Due to the small thickness $2a$ of the slabs, we are going to impose the boundary conditions for vanishing field around the metallic rods, only on $(U_1+U_2)$ specific points: the centers of the circular bounds: $E_{inc}(\rho_q(v_q), \phi_q(v_q))+E_{scat}(\rho_q(v_q), \phi_q(v_q))=0$ ($u_q=1,\cdots,U_q~,~q=1,2$). That yields to:
\begin{eqnarray}
\left[\begin{array}{cc} \textbf{M}_{11} & \textbf{M}_{12} \\ \textbf{M}_{21} & \textbf{M}_{22} \end{array}\right]\cdot 
\left[\begin{array}{cc} \textbf{k}_1 \\ \textbf{k}_2 \end{array}\right]=
\left[\begin{array}{cc} \textbf{e}_{1,inc} \\ \textbf{e}_{2,inc} \end{array}\right].
\label{eq:MetallicSystem}
\end{eqnarray}
The vectors of the surface current $\textbf{k}_q$, are containing the unknown quantities $k_q(u_q)$, while the elements of the matrices and the constant vectors are given explicitly in the expressions (\ref{eq:MetallicElements}) of the next page. The indexes take the first two positive integer values: $p,q=1,2$. Once the unknown parameters $\textbf{k}_q$ are determined, the scattered field into the vacuum background can be directly evaluated through (\ref{eq:MetallicIntegral}) and the actual induced current $K_{act}(d)$ on the metallic wedge can be found as a function of the tangential distance on the strip $d$.

\begin{figure*}[t]
\hrulefill
\setcounter{equation}{4}
\begin{subequations}
\begin{equation}
M_{qq}(v_q, u_q)=\frac{\pi a \omega \mu_0}{2}
\left\{\begin{array}{cc} J_0(k_0a)H_0^{(2)}(k_0|\rho_q(v_q)-\rho_q(u_q)|) & ,v_q\ne u_q \\ H_0^{(2)}(k_0a) & ,v_q=u_q\end{array}\right.,
\label{eq:DiagonalM}
\end{equation}
\begin{equation}
M_{pq}(v_p, u_q)=\frac{\pi a \omega \mu_0}{2}J_0(k_0a)
H_0^{(2)}\left(k_0\sqrt{\rho^2_p(v_p)+\rho^2_q(u_q)-2\rho_p(v_p)\rho_q(u_q)\cos2\theta}\right),
\label{eq:OffDiagonalM}
\end{equation}
\begin{equation}
e_{q,inc}(v_q)=E_{inc}(\rho_q(v_q), \phi_q(v_q)).
\label{eq:ConstantVectors}
\end{equation}
\label{eq:MetallicElements}
\end{subequations}
\end{figure*} 

\subsection{Dielectric Corner}
As far as the dielectric case is concerned, we employ another pivotal formula known as scattering integral \cite{ScatteringIntegral} which, in the present case, can be written as in (\ref{eq:DielectricIntegral}), shown in the next page.
\begin{figure*}[t]
\hrulefill
\setcounter{equation}{5}
\begin{equation}
E_{scat}(\rho,\phi)=k_0^2(\epsilon_r-1)
\left[\sum_{u_1=1}^{U_1}e_1(u_1)\int_{s_1(u_1)}G(\rho,\phi,P_S,\Phi_S)dS+\sum_{u_2=1}^{U_2}e_2(u_2)\int_{s_2(u_2)}G(\rho,\phi,P_S,\Phi_S)dS\right].
\label{eq:DielectricIntegral}
\end{equation}
\end{figure*}
The notation $e_q(u_q)$ has been reserved for the unknown electric field on the $u_q$-th circular disk. The surface variable $S$ is used for the area integrations on the cross sections of the pins with respect to the arguments $(P_S, \Phi_S)$. By testing (\ref{eq:DielectricIntegral}) on the centers of the consecutive dielectric cylinders, the following linear system is obtained:
\begin{eqnarray}
\left\{\textbf{I}+(\epsilon_r-1)\left[\begin{array}{cc} \textbf{D}_{11} & \textbf{D}_{12} \\ \textbf{D}_{21} & \textbf{D}_{22} \end{array}\right]\right\}\cdot 
\left[\begin{array}{cc} \textbf{e}_1 \\ \textbf{e}_2 \end{array}\right]=
\left[\begin{array}{cc} \textbf{e}_{1,inc} \\ \textbf{e}_{2,inc} \end{array}\right],
\label{eq:DielectricSystem}
\end{eqnarray}
where $\textbf{I}$ is the $(U_1+U_2)\times(U_1+U_2)$ identity matrix. The related explicit forms for the matrices of the dielectric problem are given in (\ref{eq:DielectricElements}) which are also explicitly presented in the next page.

\begin{figure*}[t]
\hrulefill
\setcounter{equation}{7}
\begin{subequations}
\begin{equation}
D_{qq}(v_q, u_q)=\frac{j\pi}{2}
\left\{\begin{array}{cc} k_0aJ_1(k_0a)H_0^{(2)}(k_0|\rho_q(v_q)-\rho_q(u_q)|) & ,v_q\ne u_q \\ k_0aH_0^{(2)}(k_0a)-\frac{2j}{\pi} & ,v_q=u_q\end{array}\right. ,\\
\label{eq:DiagonalD}
\end{equation}
\begin{equation}
D_{pq}(v_p, u_q)=\frac{j\pi}{2}k_0aJ_1(k_0a)
H_0^{(2)}\left(k_0\sqrt{\rho^2_p(v_p)+\rho^2_q(u_q)-2\rho_p(v_p)\rho_q(u_q)\cos2\theta}\right).
\label{eq:OffDiagonalD}
\end{equation}
\label{eq:DielectricElements}
\end{subequations}
\hrulefill
\end{figure*}

Similarly to the metallic case, the scattered field by the dielectric corner is found via (\ref{eq:DielectricIntegral}), based on the fields $\textbf{e}_q$ determinable from the system. The axial polarization current imitating the sources, producing the scattering field, is defined by \cite{PolarizationCurrents}: 
\begin{eqnarray}
K_{pol}(d)=\frac{ja\omega\epsilon_0}{2}(\epsilon_r-1)E(d),
\label{eq:PolarizationCurrent}
\end{eqnarray}
where $d$ is the distance on the strip and $E(d)$ the total electric field along the slab hands.

\section{Numerical Results}
\subsection{Parameter Ranges}
Prior to proceeding to the numerical simulation and commenting on the produced graphs, we present the value ranges of the input parameters are as follows. Firstly, the operating wavelength $\lambda_0$ is kept constant and equal to 1 m ($k_0=2\pi$ rad/m) throughout the numerical simulations, while the lengths of the wedge hands are mainly chosen within the interval: $\frac{\lambda_0}{10}<L_1,L_2<2\lambda_0$. In this sense, various cases of different electrical sizes and symmetries are examined, without caring about realistic operating frequencies. The angular extent of the wedge covers all possible values ($0^o<2\theta<180^o$), while the plane wave usually travels along the positive $x$ semi-axis ($\xi=0^o$) to excite the sharp edge. When it comes to the complex relative permittivity $\epsilon_r=\Re[\epsilon_r]+j\Im[\epsilon_r]$, its real part can possess extremely large positive values, while the opposite happens for the imaginary part due to the selected time dependence $(\Re[\epsilon_r]>>1, \Im[\epsilon_r]<<-1)$. The thickness of the slabs should be much smaller compared to the hand lengths, so that our assumption for thin-cylinder discretization is justified, namely $a<0.02\lambda_0$. The quantities that are of interest include the surface currents $K_{act},K_{pol}$ and the scattering field $E_{scat}$. The average relative error in the field, around a small cylinder of radius $\rho_0$ in the vicinity of the edge is expressing how well the dielectric surface imitates the metallic one and defined as follows:
\begin{eqnarray}
DE(\rho_0)=\nonumber \\
\frac{1}{2(\pi-\theta)}\int_{\theta}^{2\pi-\theta}\frac{\left|E_{M}(\rho_0,\phi)-E_{D}(\rho_0,\phi)\right|}
{\left|E_{M}(\rho_0,\phi)\right|+\left|E_{D}(\rho_0,\phi)\right|}d\phi.
\label{eq:FieldError}
\end{eqnarray}
The symbols $E_M, E_D$ are used for the electric fields in the metallic and dielectric case respectively, while the integration is carried out numerically.

\subsection{Comparisons and Validations}
One of the scopes of the present report is to examine how well a sharp PEC boundary can be materialized with use of a magnetically inert dielectric. Given the fact that we are mainly interested in the variation of the waves close to the edge, it is meaningful to compare the response of our finite metallic construction with the case of a PEC infinite wedge. Only if the field behavior in the vicinity of the sharp corner for the finite and the infinite case is similar, it is worth to study the materialization of the sharp PEC surface with least penetrable substances. It is well-known \cite{BalanisWedge}, that the solution to the plane wave scattering by an infinite PEC wedge with the same characteristics as the examined one (angular extent $2\theta$, same placement of the coordinate system as in Fig. \ref{fig:Figu1}) is given by:
\begin{eqnarray}
E_W(\rho, \phi)=\frac{2\pi}{\pi-\theta}\cdot \nonumber \\ 
\sum_{m=1}^{+\infty}j^{\mu_m}J_{\mu_m}(k_0\rho)\sin\left[\mu_m(\pi+\xi-\theta)\right]\sin\left[\mu_m(\phi-\theta)\right],
\label{eq:WedgeField}
\end{eqnarray}
with $\mu_m=\frac{m\pi}{2(\pi-\theta)}$. Similarly to (\ref{eq:FieldError}), one can define a quantity that expresses the difference between the response of the infinite and the finite PEC structure:
\begin{eqnarray}
\Delta E(\rho_0)=\nonumber \\
\frac{1}{2(\pi-\theta)}\int_{\theta}^{2\pi-\theta}\frac{\left|E_{W}(\rho_0,\phi)-E_{M}(\rho_0,\phi)\right|}
{\left|E_{W}(\rho_0,\phi)\right|+\left|E_{M}(\rho_0,\phi)\right|}d\phi.
\label{eq:WedgeError}
\end{eqnarray}
Only if $\Delta E(\rho_0)$ is relatively low, one can proceed in investigating the conditions under which a mathematical PEC boundary can be replaced by a dielectric interface. 

\begin{figure}[t]
\centering
\subfigure[]{\includegraphics[scale =0.5]{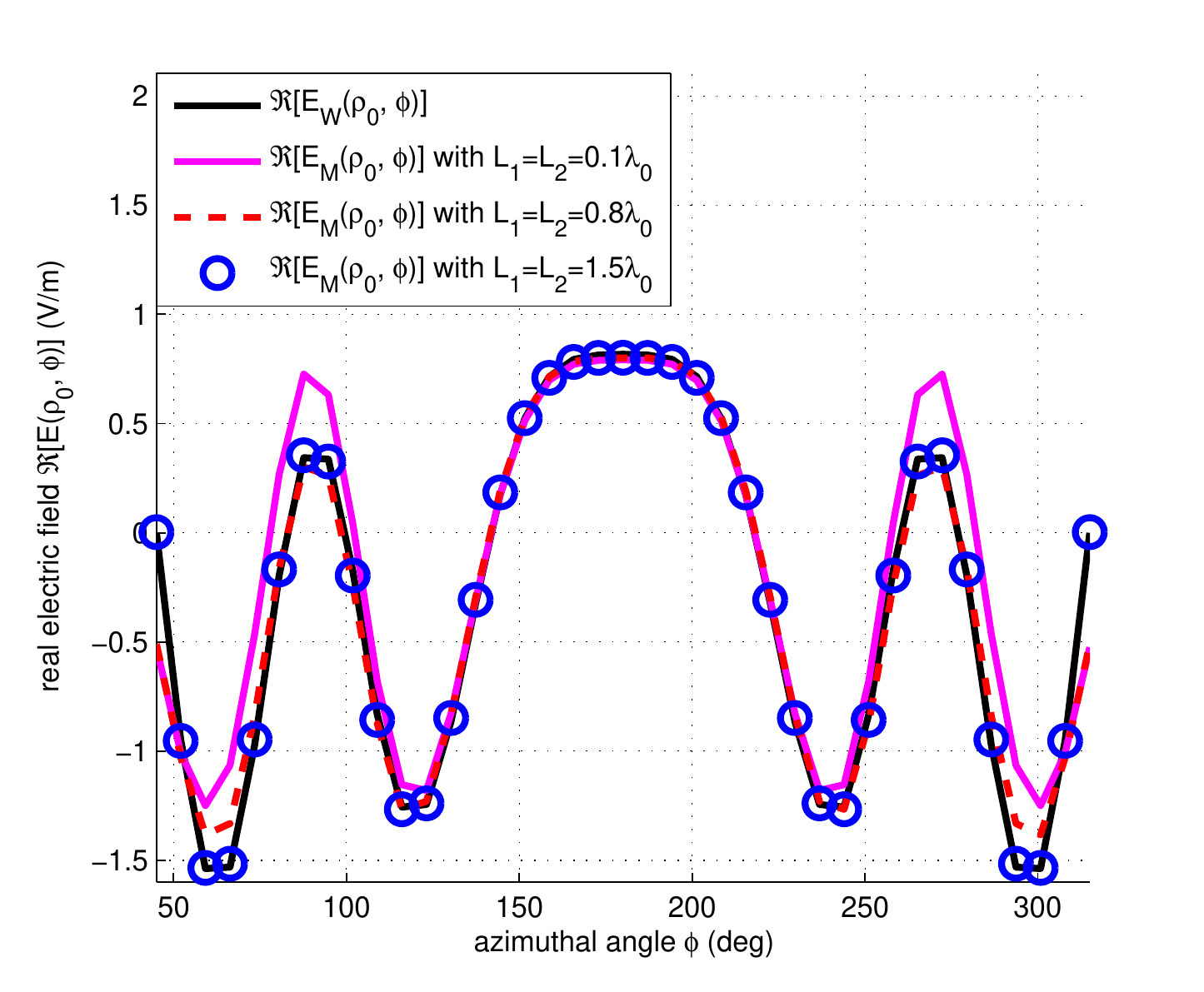}
   \label{fig:Figu8a}}
\subfigure[]{\includegraphics[scale =0.5]{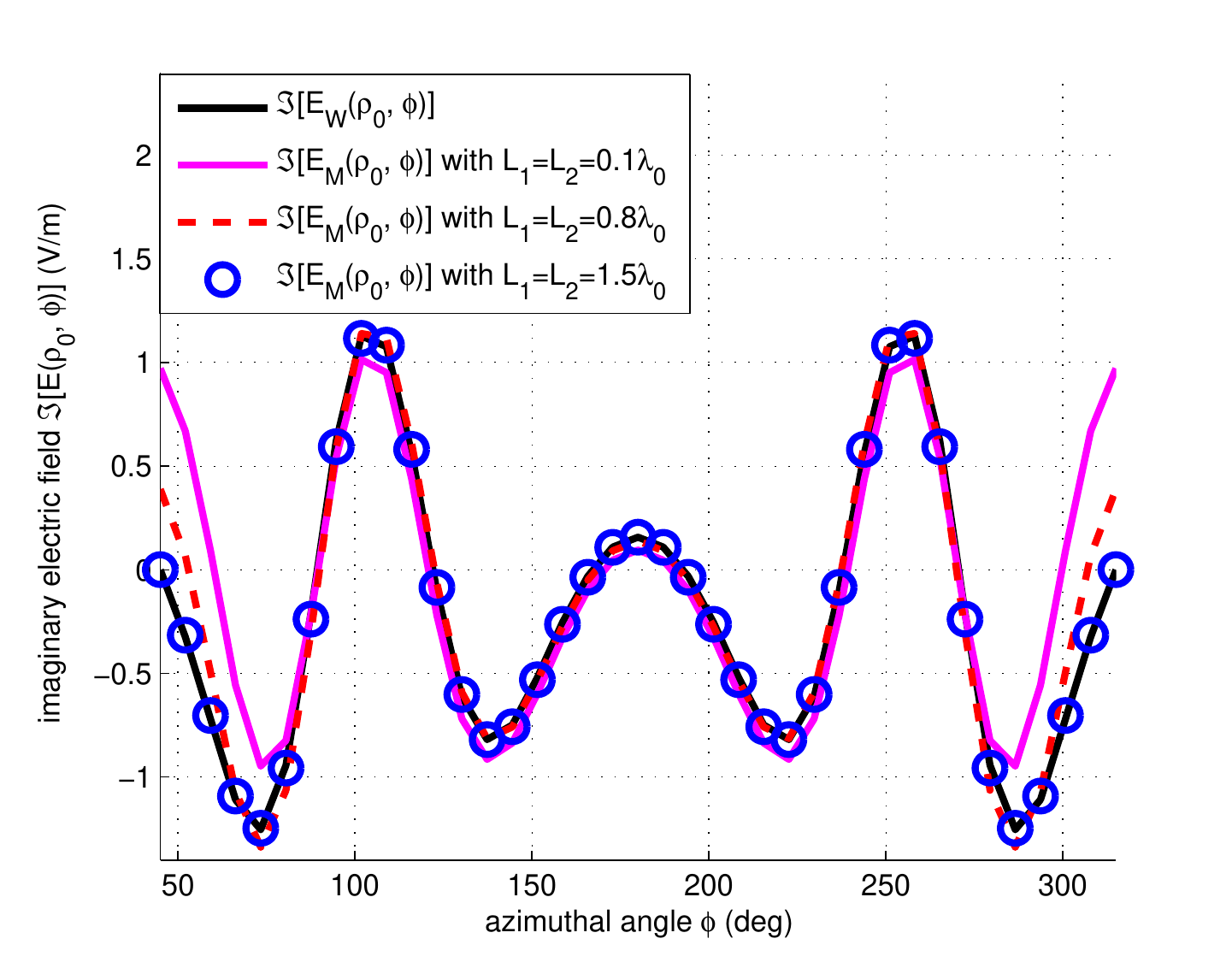}
   \label{fig:Figu8b}}
\caption{The: (a) real part and (b) imaginary part of the electric field as functions of the azimuthal angle for various lengths of the hands. Plot parameters: $\lambda_0=1$ m, $\theta=45^o$, $\xi=0^o$, $a/\lambda_0=0.015$, $\rho_0=1$ m.}
\label{fig:Figs8}
\end{figure}

\begin{figure}[t]
\centering
\subfigure[]{\includegraphics[scale =0.5]{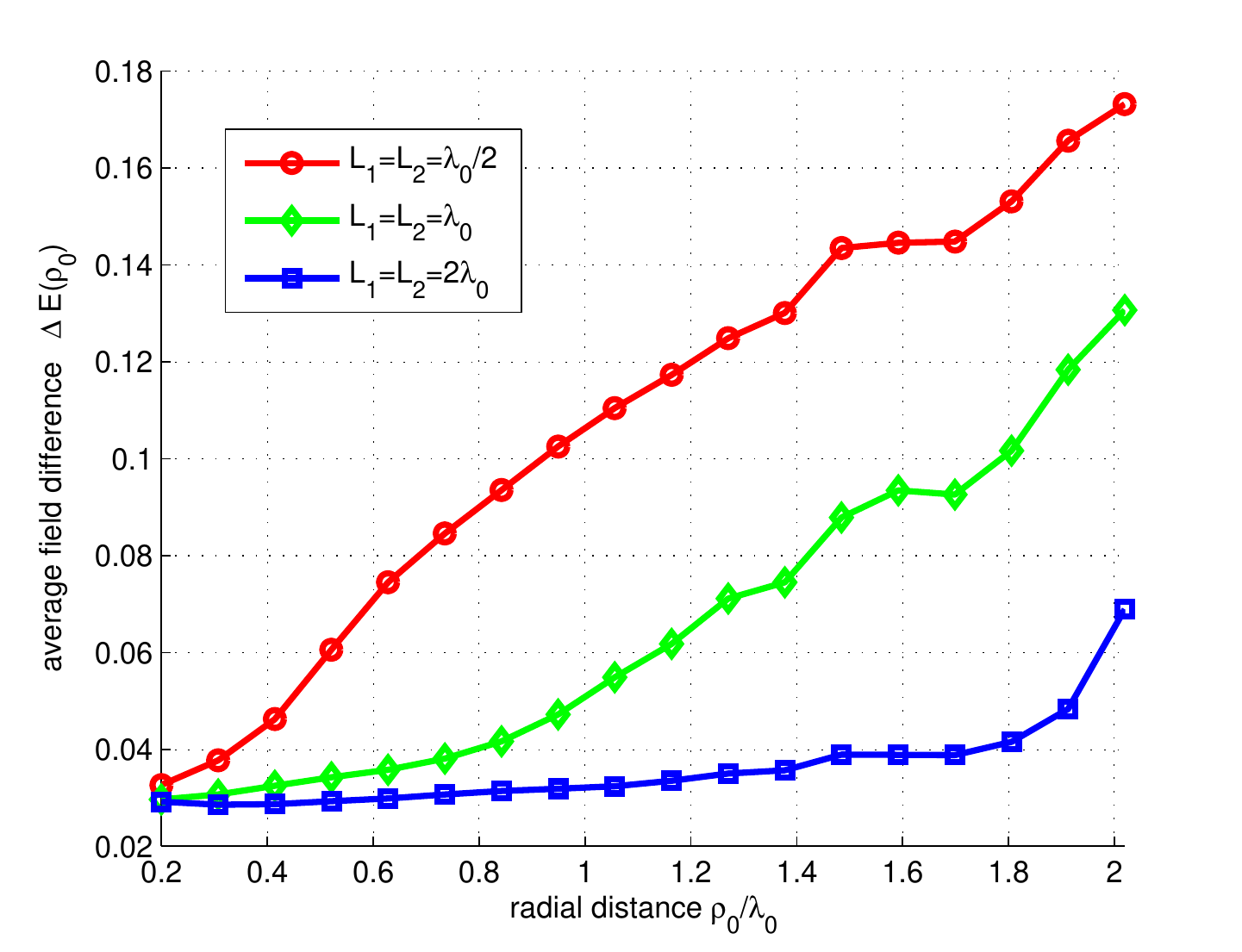}
   \label{fig:Figu9a}}
\subfigure[]{\includegraphics[scale =0.5]{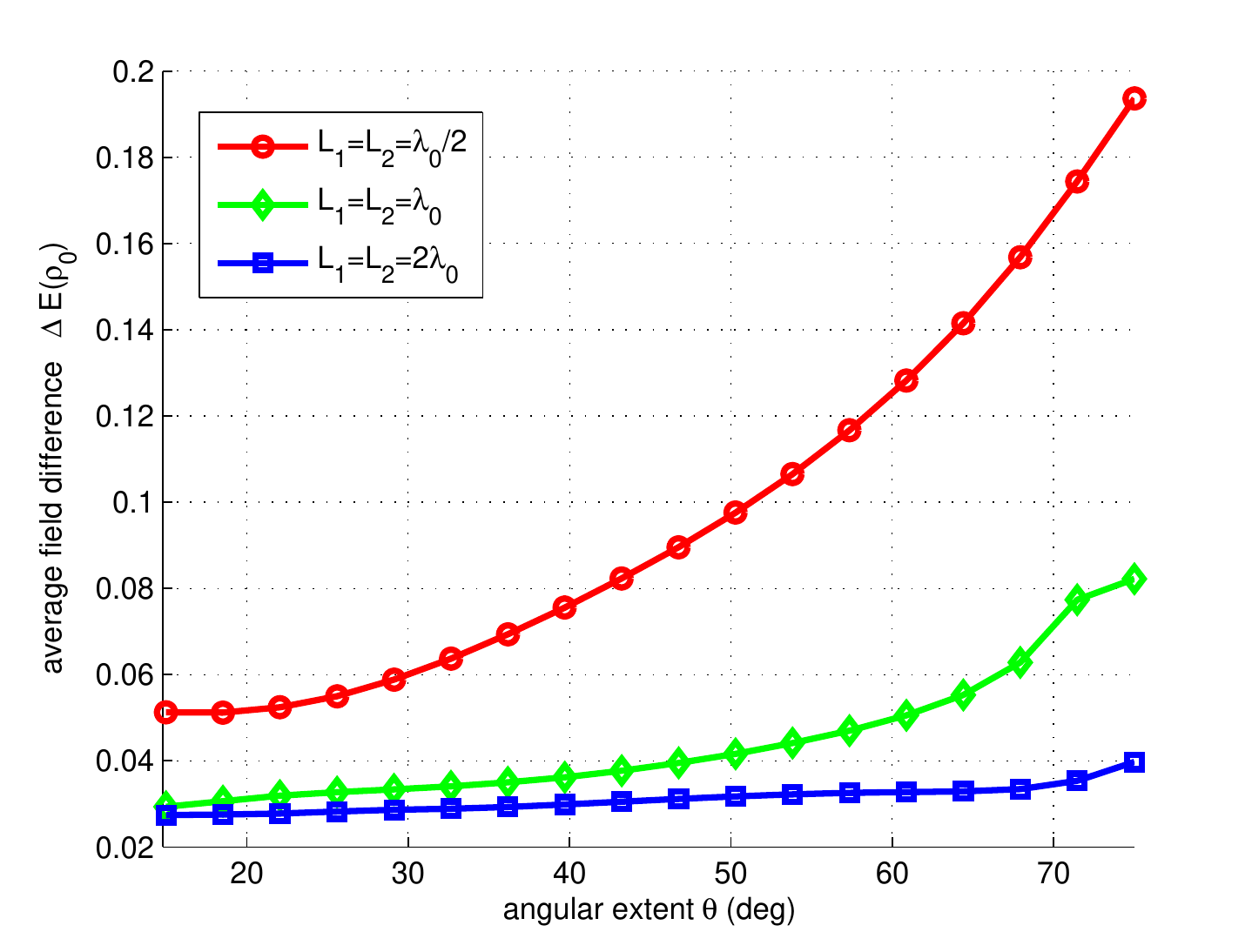}
   \label{fig:Figu9b}}
\caption{The average field difference compared with the infinite metallic edge structure as function of: (a) the radial observation distance ($\theta=45^o$) and (b) the angular extent of the wedge for various lengths of the wedge hands ($\rho_0=0.75\lambda_0$). Plot parameters: $\lambda_0=1$ m, $\xi=0^o$, $a/\lambda_0=0.015$.}
\label{fig:Figs9}
\end{figure}

In Figs \ref{fig:Figs8}, we depict the real and the imaginary part of the electric field as functions of the azimuthal angle $\phi\in(\theta,2\pi-\theta)$ for metallic corners of different $L_1=L_2$. It is natural that for larger hand lengths the curves of $E_M(\rho_0, \phi)$ gets closer to the solid black line given by $E_W(\rho_0,\phi)$. It is worth noting that when the size of the segments equals a couple of free space wavelengths, both parts of the electric field are almost identical to those of (\ref{eq:WedgeField}). In Fig. \ref{fig:Figu9a}, the difference $\Delta E(\rho_0)$ is represented as function of the electrical distance $\rho_0/\lambda_0$ for several lengths $L_1=L_2$. Again, one notices that the larger are the hands of a finite wedge, the closer is the exhibited behavior to that of an infinite construction. Moreover, the curves are upward sloping with respect to the radial observation distance which remarks that the field in the vicinity of the sharp edge is not affected by the size of the corner. In Fig. \ref{fig:Figu9b}, the difference in the field between the finite and the infinite case is shown with respect to the half angular extent $\theta$. It is noteworthy that the more oblique is the corner, the smaller is the recorded error, a feature which is justified by the more powerful response of the edge. As a conclusion, one could point out that the field near the edge ($\rho_0<<L_1,L_2$) of an infinite PEC wedge is almost equal to that of a finite wedge provided that its hands are electrically large ($L_1,L_2>\lambda_0$).

\begin{figure}[t]
\centering
\subfigure[]{\includegraphics[scale =0.5]{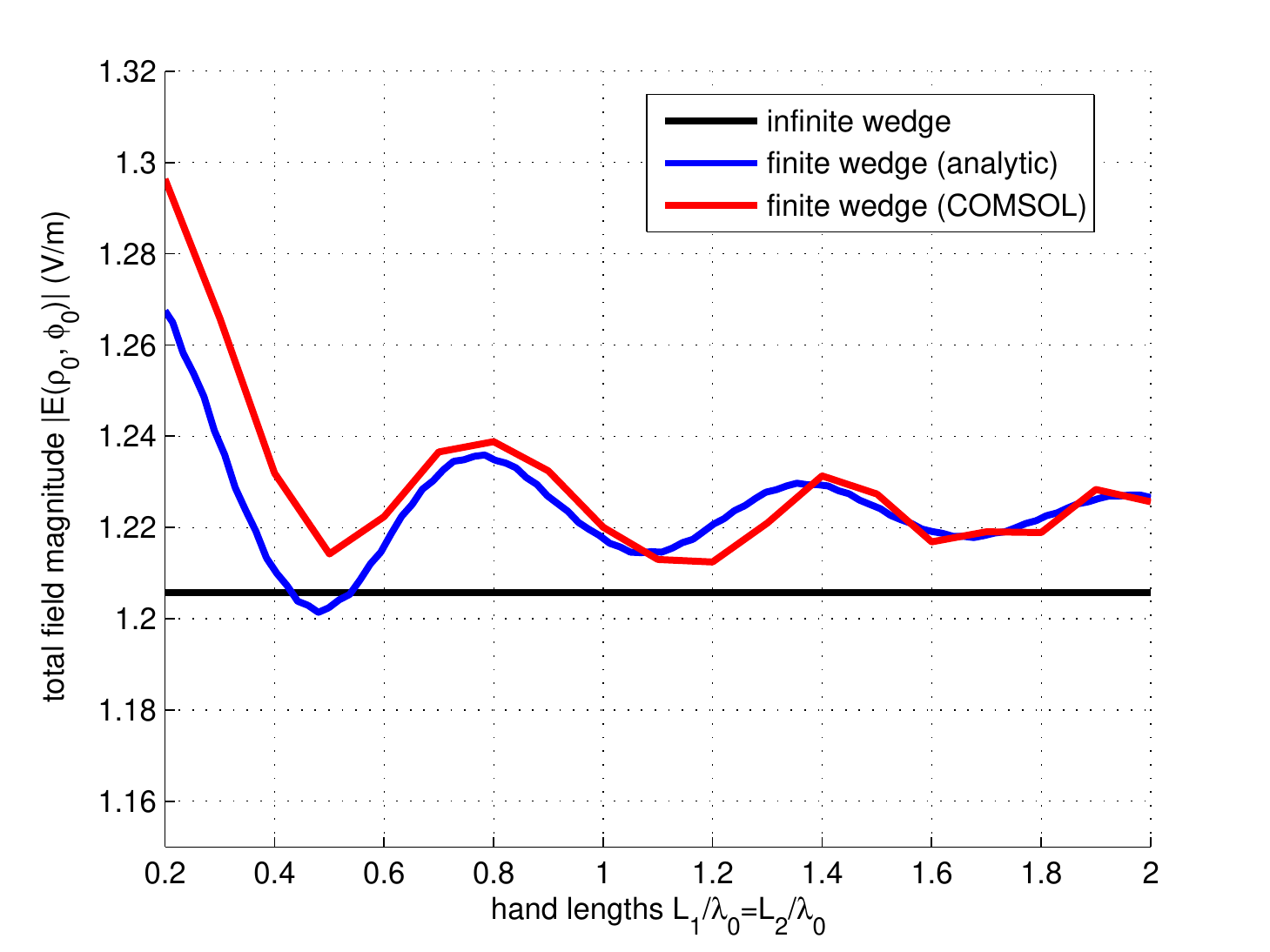}
   \label{fig:Figu10a}}
\subfigure[]{\includegraphics[scale =0.5]{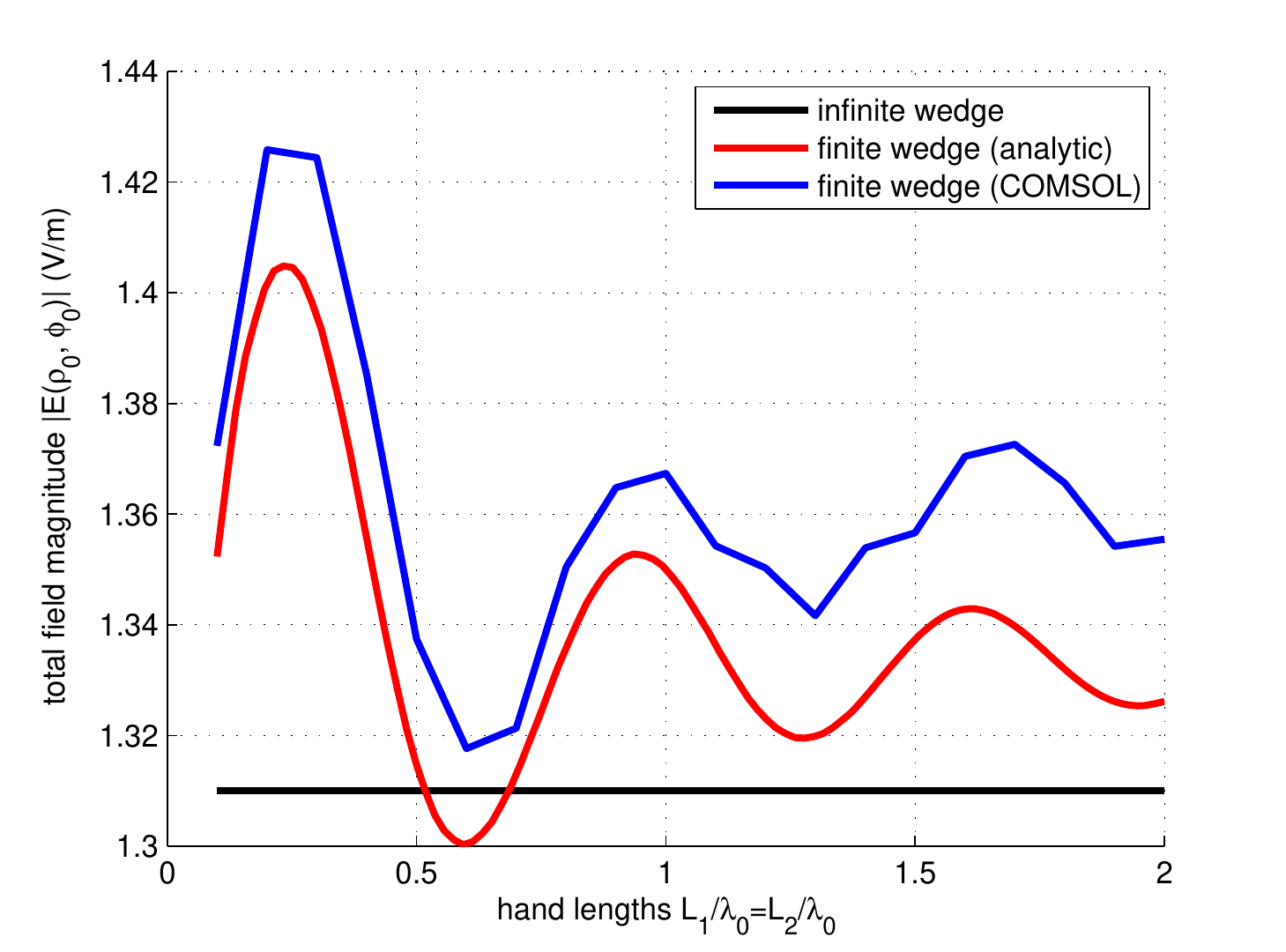}
   \label{fig:Figu10b}}
\caption{The magnitude of the total electric field at a  specific point as function of the electrical length of the two equal hands for: (a) $\theta=45^o$ and (b) $\theta=60^o$. Plot parameters: $\lambda_0=1$ m, $\xi=0^o$, $a/\lambda_0=0.015$, $\phi_0=180^o$, $\rho_0/\lambda_0=0.3$.}
\label{fig:Figs10}
\end{figure}

To provide an additional validation check for our techniques and computations, we simulated the metallic structure and the excitation depicted in Fig. \ref{fig:Figu1} using COMSOL Multiphysics commercial software \cite{COMSOL}. The magnitude of the developed (total) electric field at a specific point $(\rho, \phi)=(\rho_0, \phi_0)$ is represented as function of the electrical lengths $L_1/\lambda_0=L_2/\lambda_0$ of the hands which are of equal size. The simulation results are compared against those derived from expressions (\ref{eq:MetallicIntegral}) and (\ref{eq:WedgeField}) for the finite and the infinite wedge respectively. In Fig. \ref{fig:Figu10a}, the angular extent is taken equal to $\theta=45^o$; it is clear that the coincidence between the two curves (blue for semi-analytic approach, red for COMSOL numerical method) describing the finite wedge is quite satisfying. Naturally, the field solution for the infinite wedge (black curve) is constant and not dependent on the length of the hands. Mind that when the lengths $L_1=L_2$ are increasing, the difference between the field of the finite and the infinite structure gets diminished. In Fig. \ref{fig:Figu10b}, where the angle of the wedge is larger ($\theta=60^o$), the agreement between the COMSOL simulation and the analytical evaluation is also remarkable.

\subsection{Currents and Fields}
\begin{figure}[t]
\centering
\subfigure[]{\includegraphics[scale =0.5]{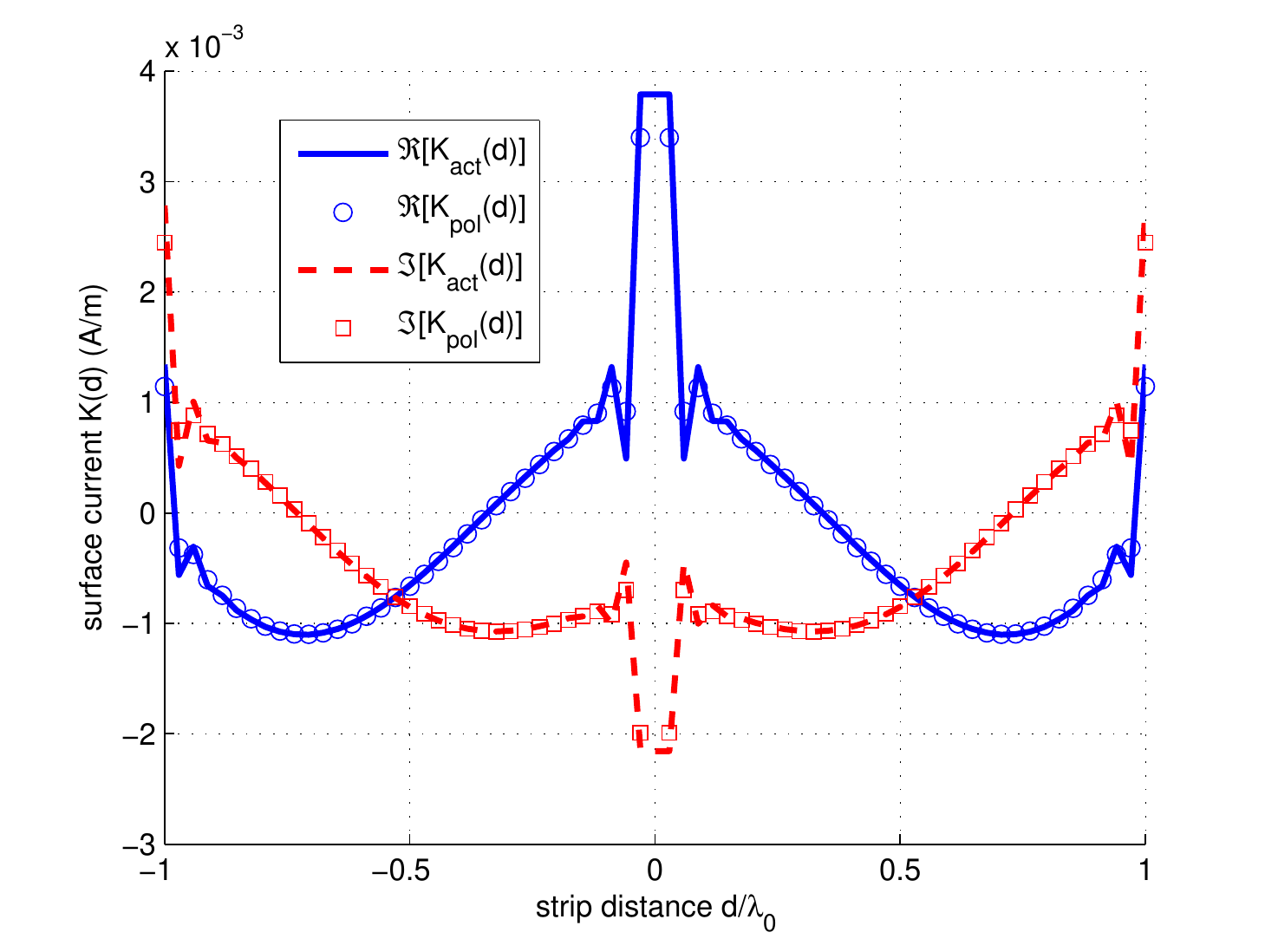}
   \label{fig:Figu2a}}
\subfigure[]{\includegraphics[scale =0.5]{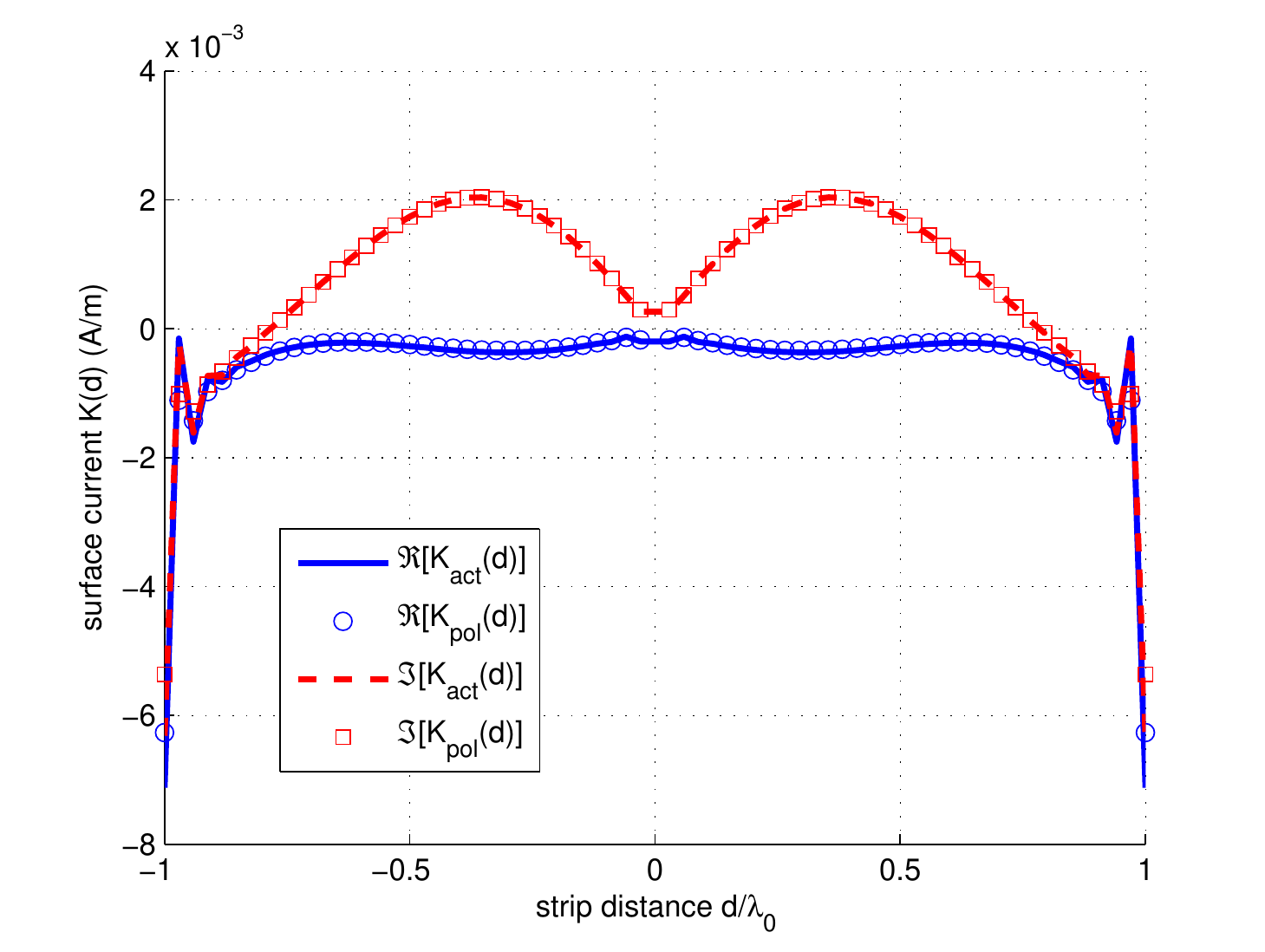}
   \label{fig:Figu2b}}
\caption{The actual and the polarization current on the metallic and the dielectric obstacle respectively, as functions of the tangential distance on the strip for: (a) $\xi=0^o$ and (b) $\xi=180^o$. Plot parameters: $\lambda_0=1$ m, $L_1/\lambda_0=1$, $L_2/\lambda_0=1$, $\epsilon_r=800$, $\theta=45^o$, $a/\lambda_0=0.015$.}
\label{fig:Figs2}
\end{figure}

In Figs \ref{fig:Figs2}, we show the variation of the actual current $K_{act}$ induced on the metallic wedge and the respective polarization current $K_{pol}$, when $\epsilon_r=800$, as function of the normalized tangential distance on the strip $d/\lambda_0$ whose negative values correspond to the lower hand of the wedge. The symmetry of the graphs is dictated by the related symmetry of the structure ($L_1=L_2=\lambda_0$). Regardless of the incidence angle ($\xi=0^o$ in Fig. \ref{fig:Figu2a} and $\xi=180^o$ in Fig. \ref{fig:Figu2a}), there is a remarkable coincidence between the two quantities which shows that the limit of the polarization current for $\Re[\epsilon_r]\rightarrow+\infty$, equals the actual current on the PEC surface. When the structure is illuminated from the left side ($\xi=0^o$), a maximum current is recorded on the corner ($d/\lambda_0=0$), while in the case of right-sided incidence ($\xi=180^o$), the quantity vanishes at the same point. Note also the substantial charge concentration along the ending edges of the two strips ($d/\lambda_0=\pm1$), is considerable for both excitation waves.

\begin{figure}[t]
\centering
\subfigure[]{\includegraphics[scale =0.5]{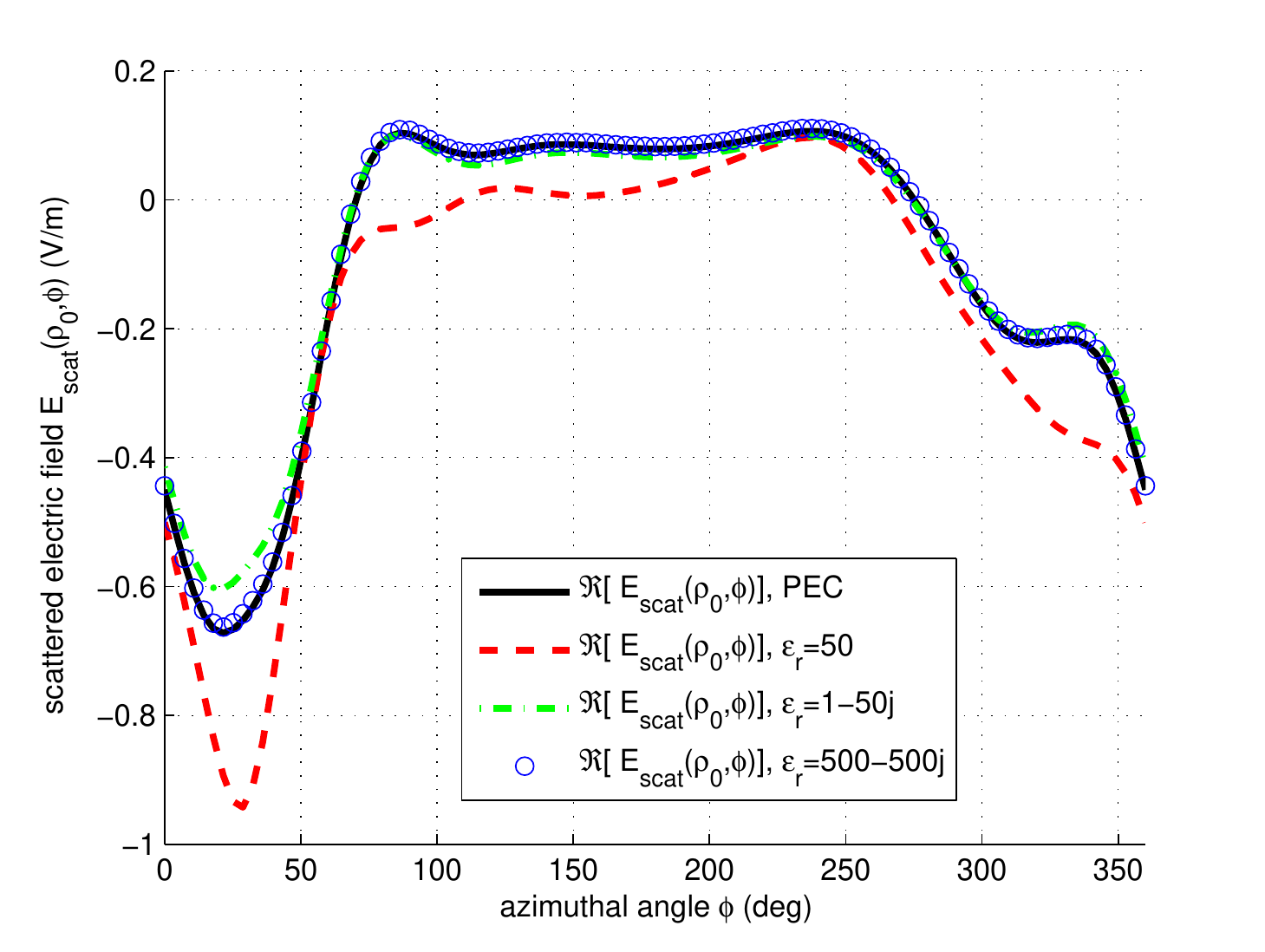}
   \label{fig:Figu3a}}
\subfigure[]{\includegraphics[scale =0.5]{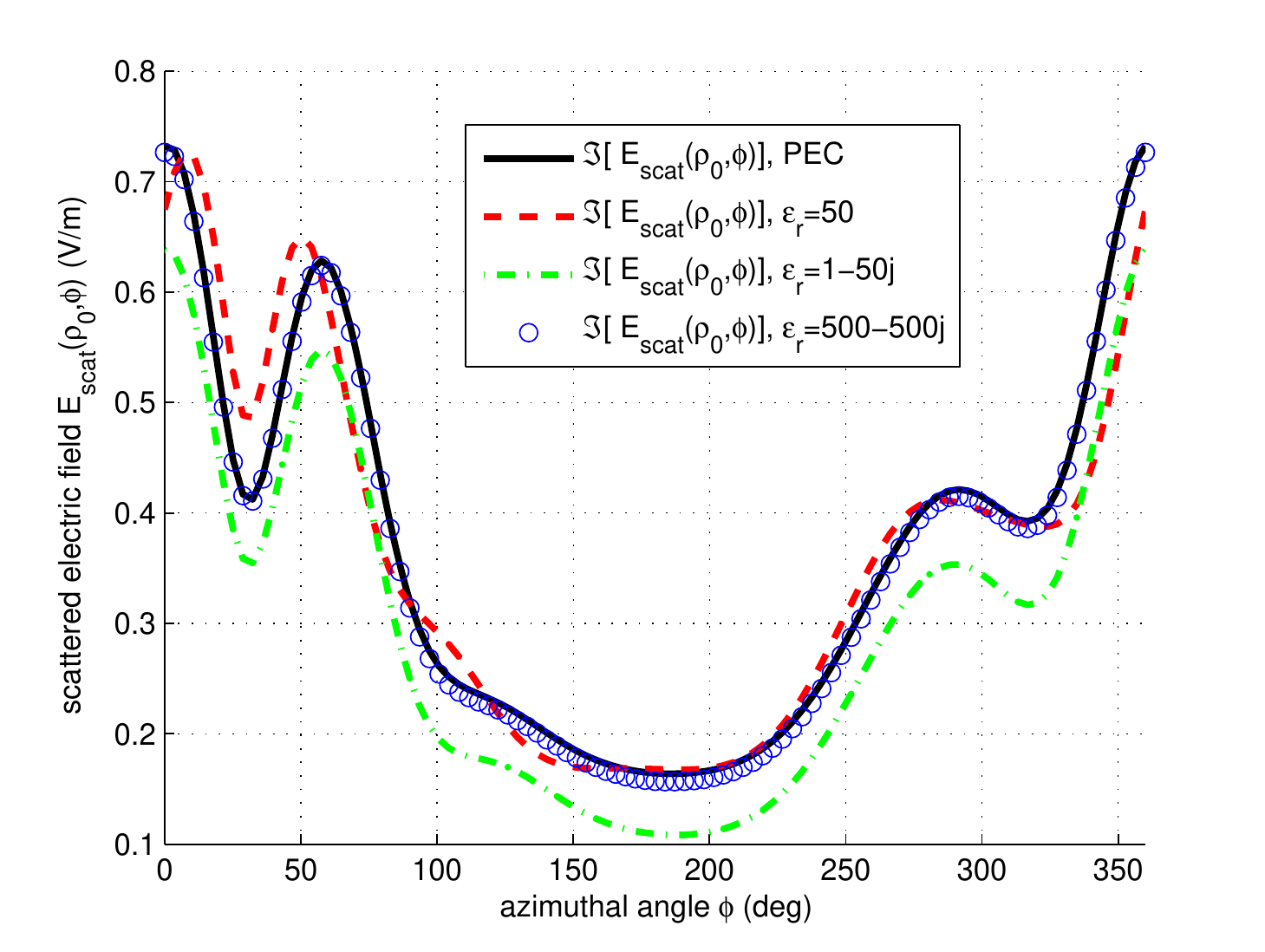}
   \label{fig:Figu3b}}
\caption{The distribution of the: (a) real part and (b) imaginary part of the scattered electric field as functions of the azimuthal angle for various relative complex permittivities of the dielectric. Plot parameters: $\lambda_0=1$ m, $L_1=1$ m, $L_2/\lambda_0=0.5$, $\theta=30^o$, $\xi=0^o$, $a/\lambda_0=0.015$, $\rho_0/\lambda_0=1.2$.}
\label{fig:Figs3}
\end{figure}

In Figs \ref{fig:Figs3}, the real and imaginary parts of the scattering electric component, are represented around a circular bound including the whole structure, as functions of the azimuthal angle $\phi$, for the metallic strip and various dielectric slabs of different permittivities. Naturally, the best fitting to the PEC case (solid black line) is achieved when both $\Re[\epsilon_r],\Im[\epsilon_r]$ are extremely high (blue dots). Nevertheless, when only one part of the complex permittivity is substantial, the coincidence of the curves is more satisfying for the opposite part of the scattering field. In particular, the red dashed line in Fig. \ref{fig:Figu3a} differs much from the black solid one compared to the variation of the green dashed one. The opposite behavior is exhibited in Fig. \ref{fig:Figu3b}.

\subsection{Average Errors}
\begin{figure}[t]
\centering
\subfigure[]{\includegraphics[scale =0.5]{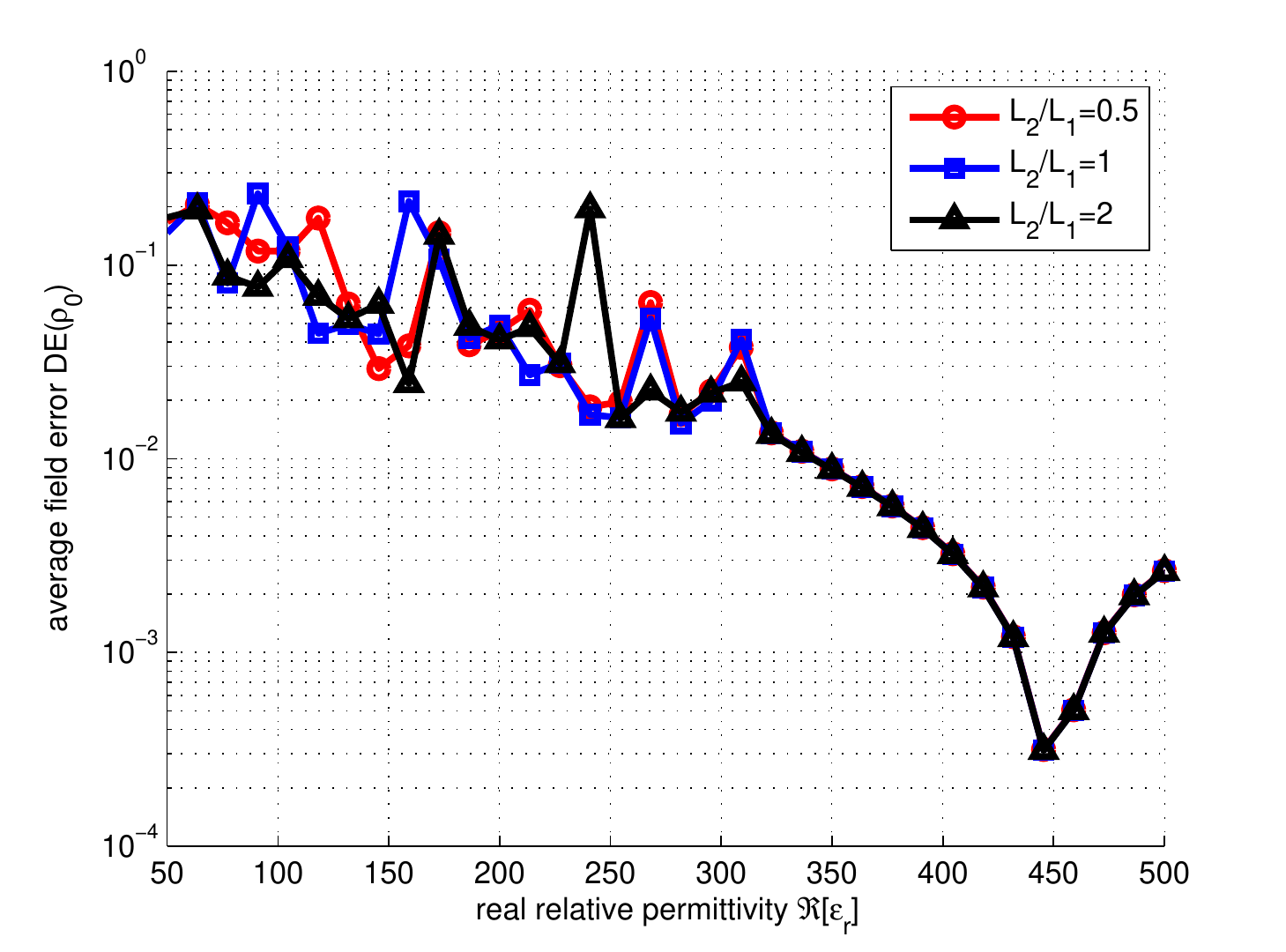}
   \label{fig:Figu4a}}
\subfigure[]{\includegraphics[scale =0.5]{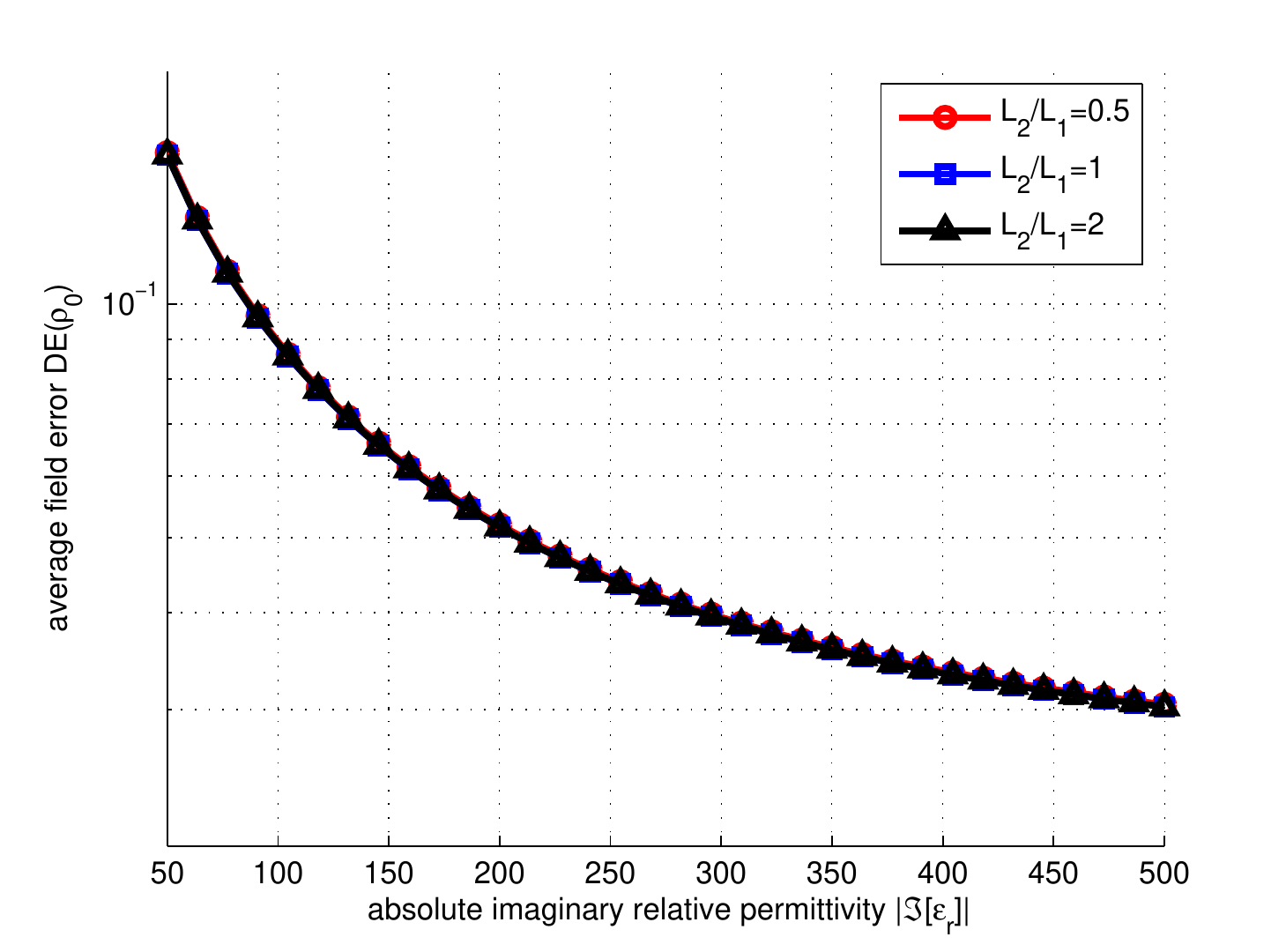}
   \label{fig:Figu4b}}
\caption{The average relative error of the electric field around the edge as function of: (a) the real part of the relative complex permittivity of the dielectric and (b) the absolute imaginary part of the relative complex permittivity of the dielectric, for various lengths of the lower hand. Plot parameters: $\lambda_0=1$ m, $L_1/\lambda_0=1$, $\theta=30^o$, $\xi=0^o$, $a/\lambda_0=0.015$, $\rho_0/\lambda_0=0.1$.}
\label{fig:Figs4}
\end{figure}

In Figs \ref{fig:Figs4}, the average relative error $DE(\rho_0)$ is depicted (in logarithmic plot) for several lengths of the strip hands, with respect to the real and the imaginary part of the permittivity $\epsilon_r$. The error gets much more diminished for larger real parts than for more substantial imaginary parts. In other words, it is preferable to model a PEC wedge with a strong dielectric instead of using an equally strong conductor. Note also the resonance oscillations  of Fig. \ref{fig:Figu4a}, recorded for somehow small $\Re[\epsilon_r]$, in contrast to the smoothly decaying curves of Fig. \ref{fig:Figu4b}. Mind that the three curves in Fig. \ref{fig:Figu4b} (and partially in Fig. \ref{fig:Figu4a}), are almost identical each other; this indicates that the lengths of the wedges's hands do not play a significant role in the behavior of the device which is dominated by the presence of the sharp corner. However, the most fascinating feature of these results concerns the minimum of Fig. \ref{fig:Figu4a} observed close to $\Re[\epsilon_r]=450$. It seems rather peculiar, given the fact that the larger permittivity is used, the smaller the error should be measured. Such a seeming contradiction is justified by the finite thickness of the wedge; in fact, there is a critical point of $\Re[\epsilon_r]$, above which the electrical thickness of the strip gets stronger, namely the transmission mechanism is activated and gradually prevails over the competing reflection mechanism. In other words, the reflection is closer to 100\% (characteristic of the metallic surfaces) when the slab is both electrically dense (large $\Re[\epsilon_r]$) and electrically thin (small $2k_0a\sqrt{\Re[\epsilon_r]}$), otherwise it behaves more as a layer than a strip and therefore the transmission through itself is not negligible (``large-$a$ limit'').

\begin{figure}[t]
\centering
\subfigure[]{\includegraphics[scale =0.5]{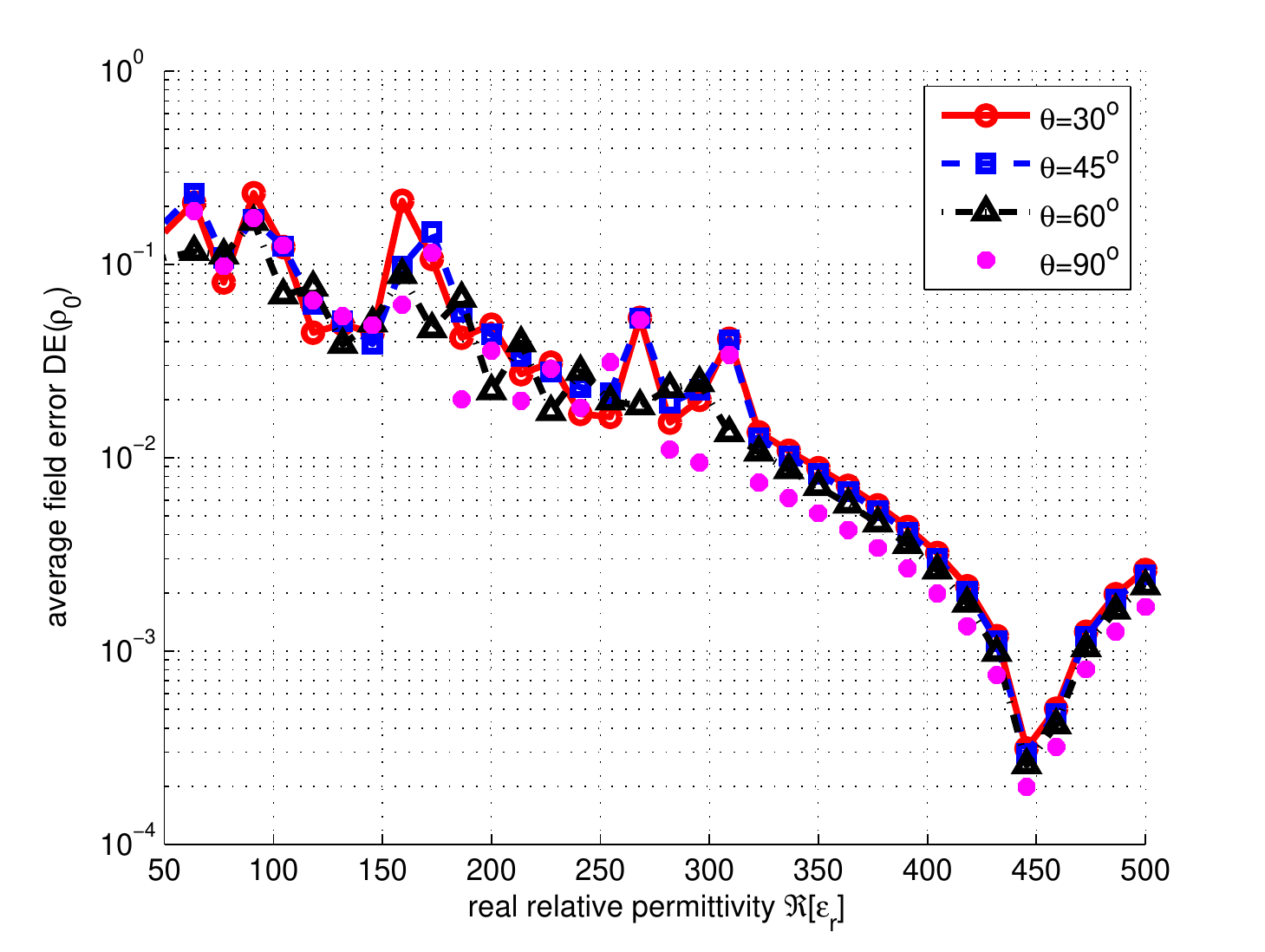}
   \label{fig:Figu5a}}
\subfigure[]{\includegraphics[scale =0.5]{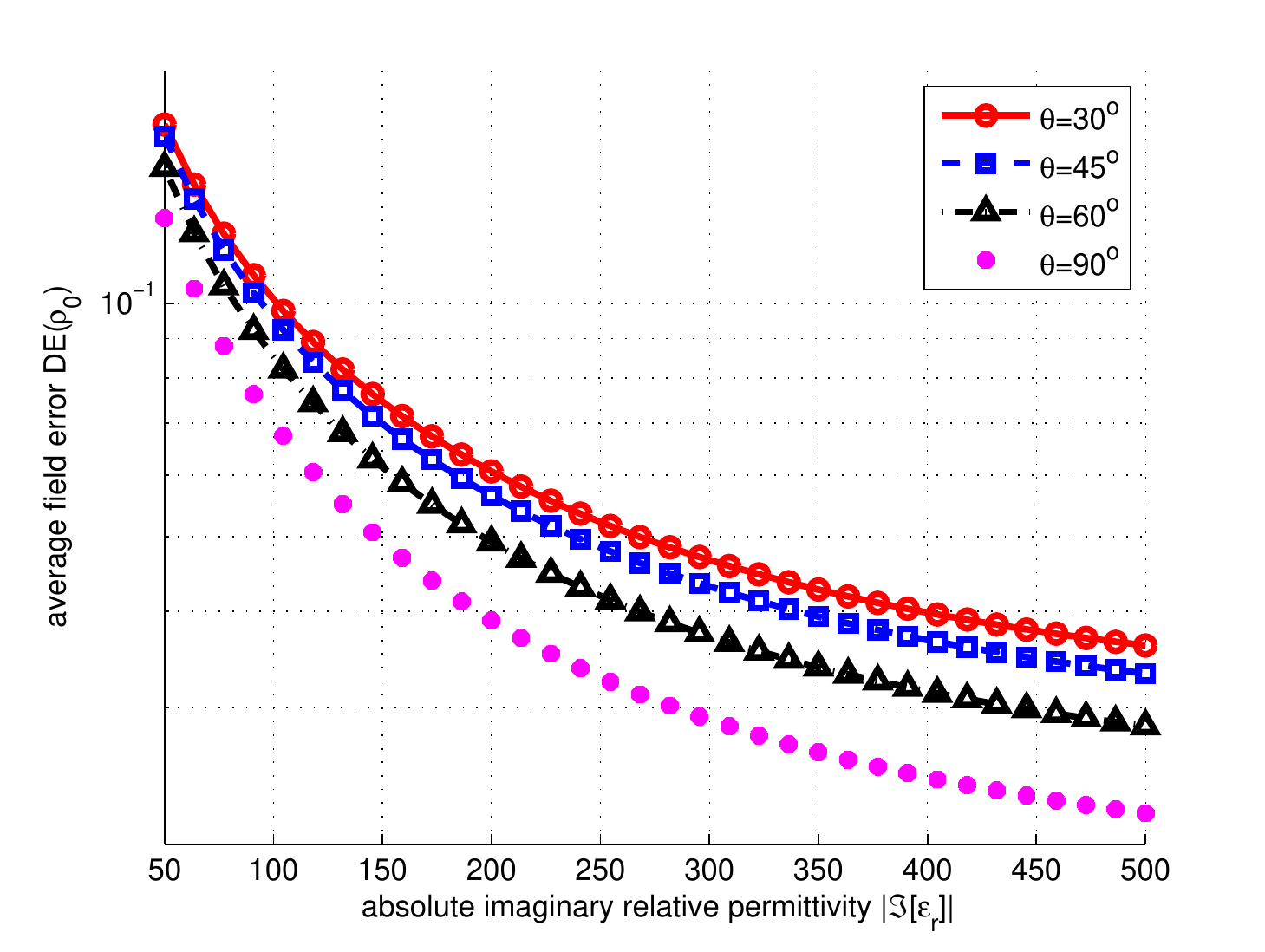}
   \label{fig:Figu5b}}
\caption{The average relative error of the electric field around the edge as function of: (a) the real part of the relative complex permittivity of the dielectric and (b) the absolute imaginary part of the relative complex permittivity of the dielectric, for various angular extent of the corner. Plot parameters: $\lambda_0=1$ m, $L_1/\lambda_0=1$, $L_2/\lambda_0=1$, $\xi=0^o$, $a/\lambda_0=0.015$, $\rho_0/\lambda_0=0.1$.}
\label{fig:Figs5}
\end{figure}

In Figs \ref{fig:Figs5}, the average relative error $DE(\rho_0)$ is shown as function of both parts of $\epsilon_r$, for various angular extents of the considered wedge. The results are similar to those of Figs \ref{fig:Figs4}, as the aforementioned minimum (``large-$a$ limit''), where the relay from reflection to transmission mechanism happens, is also noticed in Fig. \ref{fig:Figu5a}. Nonetheless, the curves in Fig. \ref{fig:Figu5b} are more dispersive compared to the the respective of Fig. \ref{fig:Figu4b}, a feature that remarks the effect of the corner angle (much more than this of the hand lengths) on the response of the regarded structure. Again the average errors for increasing real part in Fig. \ref{fig:Figu4a} are lower than those of Fig. \ref{fig:Figu4b}, a feature that is attributed to thermal losses induced by the nonzero material conductivity.

\begin{figure}[t]
\centering
\subfigure[]{\includegraphics[scale =0.5]{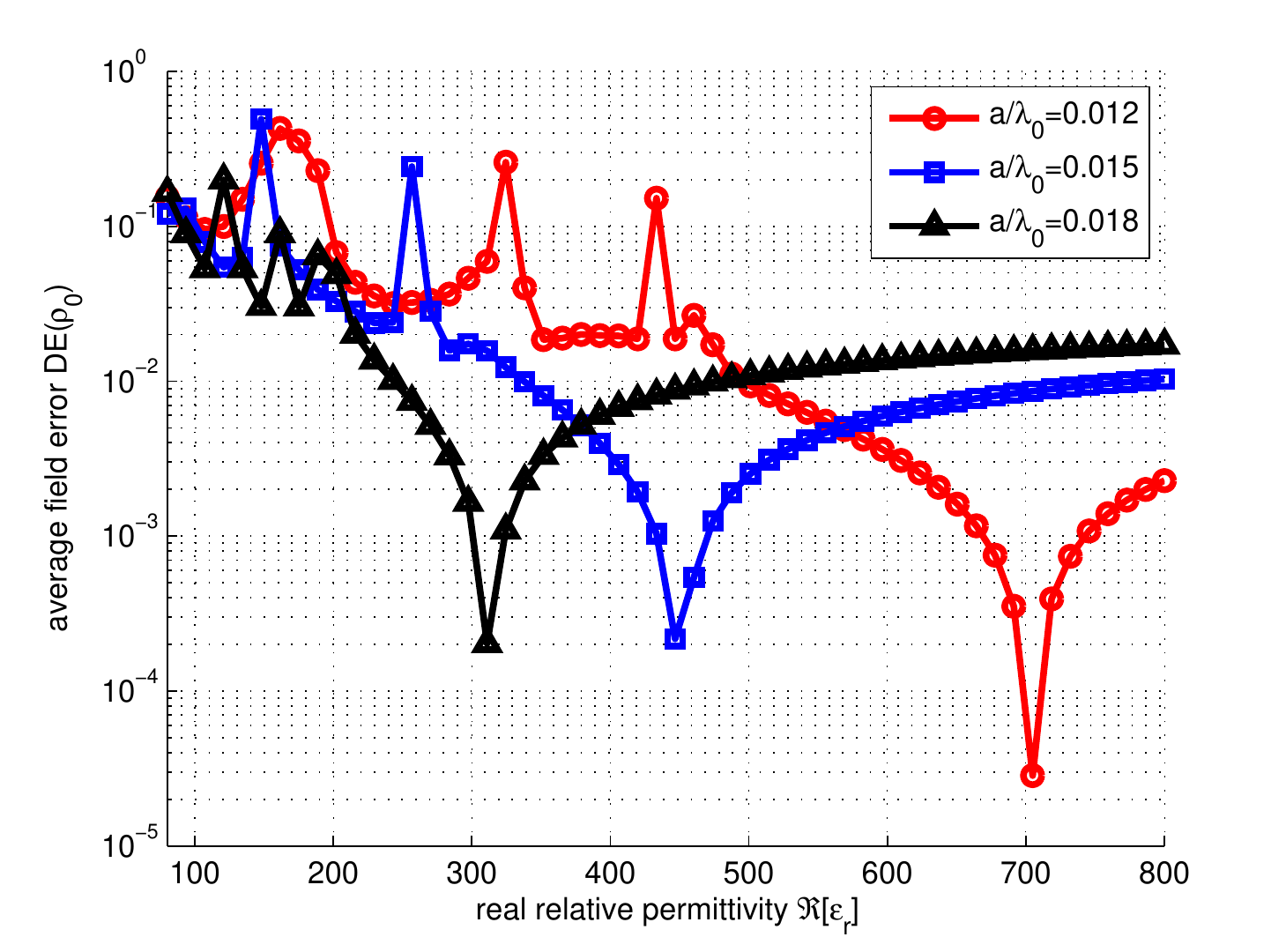}
   \label{fig:Figu6a}}
\subfigure[]{\includegraphics[scale =0.5]{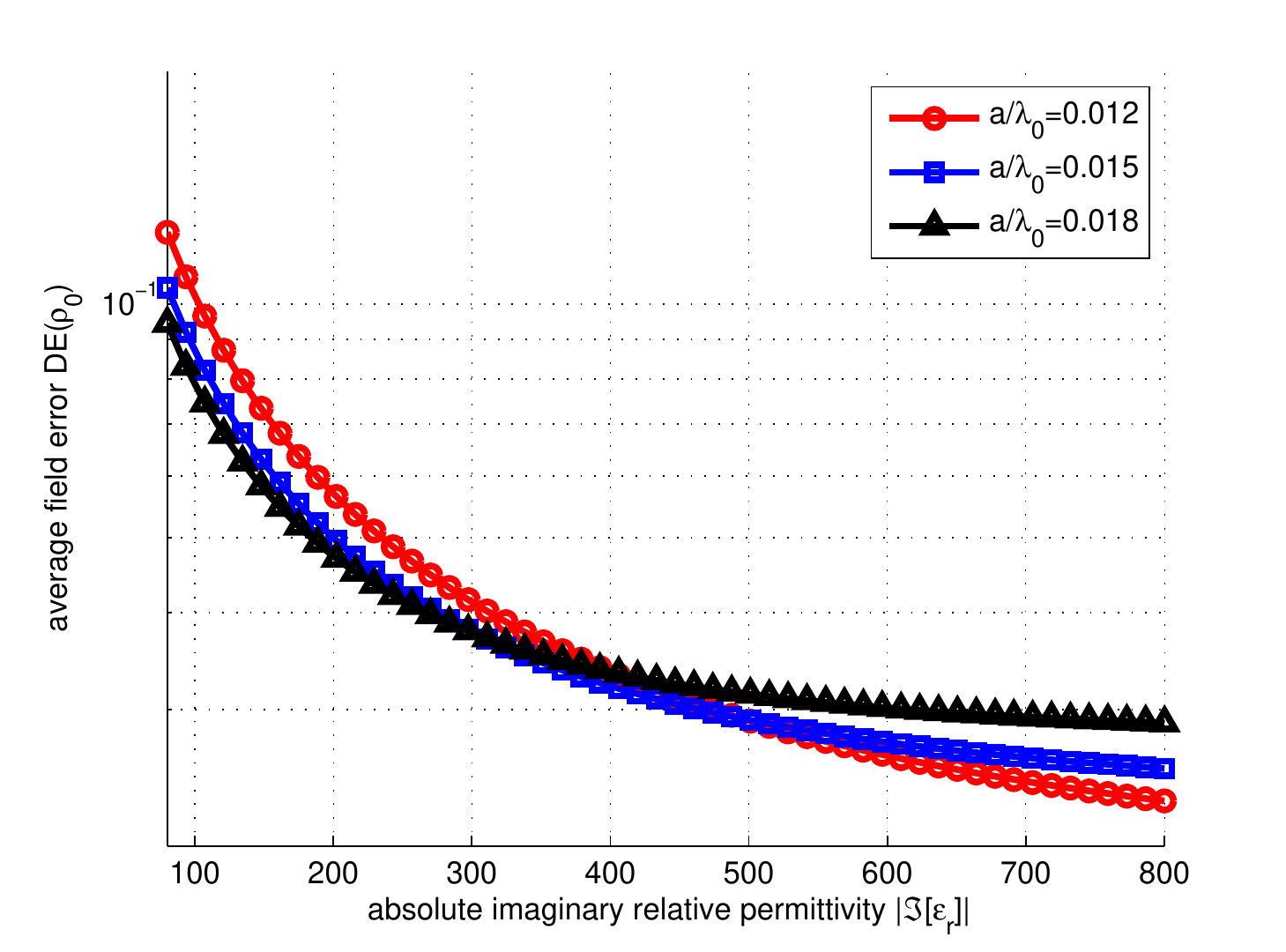}
   \label{fig:Figu6b}}
\caption{The average relative error of the electric field around the edge as function of: (a) the real part of the relative complex permittivity of the dielectric and (b) the absolute imaginary part of the relative complex permittivity of the dielectric, for various thicknesses of the strip. Plot parameters: $\lambda_0=1$ m, $L_1/\lambda_0=1$, $L_2/\lambda_0=1$, $\theta=45^o$, $\xi=0^o$, $\rho_0/\lambda_0=0.1$.}
\label{fig:Figs6}
\end{figure}

In Figs \ref{fig:Figs6}, the error quantity $DE(\rho_0)$ is represented with respect to real and imaginary parts of $\epsilon_r$ for several thicknesses $a/\lambda_0$ of the strips. The described transition from one mechanism to the other for varying $\Re[\epsilon_r]$, is clearly shown in Fig. \ref{fig:Figu6a}, where the curves reach their minimum levels at larger $\Re[\epsilon_r]$ for thinner slabs. In Fig. \ref{fig:Figu6b}, one observes that when the thickness is small, better modeling is achieved with larger $|\Im[\epsilon_r]|$, when the structure behaves as a thin conducting strip. On the contrary, more accurate results are recorded when considering thicker slabs with small-$|\Im[\epsilon_r]|$ dielectrics when the structure reacts chiefly as a rectangular scatterer.

\section{Conclusions}
An attempt to imitate the reaction of a sharp metallic corner to an incident wave with use of magnetically inert materials, has been performed in this study. A semi-analytic technique is developed to compare the scattering field in both cases. Several graphs of the field difference between the metallic and the dielectric case with respect to the complex permittivity of the constituent material and the geometrical characteristics of the structure, are presented and interpreted.

An interesting expansion of the described work would be to use magnetically active and anisotropic materials to model sharp metallic boundaries. Additionally, different excitations could be considered in order to test the robustness of the proposed PEC modeling with respect to different types of electromagnetic interaction. Finally, similar approaches would be useful in modeling other types of boundaries such as perfect electromagnetic wedges or sharp impedance surfaces.

\end{document}